\documentclass[%
reprint,
superscriptaddress,
amsmath,amssymb,
aps, pra, 
floatfix,
nobalancelastpage,
longbibliography
]{revtex4-2}

\usepackage{cancel}
\usepackage{graphicx}
\usepackage{bm}
\usepackage{textgreek}
\usepackage[mathlines]{lineno}
\usepackage{booktabs}
\usepackage{wasysym}
\usepackage{multirow} 
\usepackage{hyperref}      
\usepackage[capitalize, nameinlink]{cleveref} 
\usepackage{placeins}
\usepackage{siunitx}
\usepackage{verbatim}

\newcommand{\getTitle}{Perturbative sensing of nanoscale materials with millimeter-wave photonic crystals}

\hypersetup{
    colorlinks=true,
    linkcolor=blue,   
    urlcolor=cyan,
    citecolor=blue,
    pdftitle={\getTitle},
    pdfpagemode=FullScreen,
}

\newcommand{\nocontentsline}[3]{}
\let\origcontentsline\addcontentsline
\newcommand\stoptoc{\let\addcontentsline\nocontentsline}
\newcommand\resumetoc{\let\addcontentsline\origcontentsline}

\begin{document}
\title{\getTitle}

\author{Kevin K. S. Multani} 
\thanks{These authors contributed equally to this work.}
\affiliation{Department of Physics, Stanford University}
\affiliation{Department of Applied Physics \& E.L. Ginzton Laboratory, Stanford University}
\affiliation{SLAC National Laboratory, Menlo Park}
\author{Zhurun Ji}
\thanks{These authors contributed equally to this work.}
\affiliation{Department of Physics, Stanford University}
\affiliation{SLAC National Laboratory, Menlo Park}
\author{Wentao Jiang}
\affiliation{Department of Applied Physics \& E.L. Ginzton Laboratory, Stanford University}
\author{Siyuan Qi}
\affiliation{Department of Physics, Stanford University}
\author{Akasha G. Hayden}
\affiliation{Department of Electrical Engineering \& E.L. Ginzton Laboratory, Stanford University}
\author{Gitanjali Multani}
\affiliation{Department of Applied Physics \& E.L. Ginzton Laboratory, Stanford University}
\author{Sharon Ruth S. Platt}
\affiliation{Department of Applied Physics \& E.L. Ginzton Laboratory, Stanford University}
\author{Emilio A. Nanni}
\affiliation{SLAC National Laboratory, Menlo Park}
\author{Zhi-Xun Shen}
\affiliation{Department of Physics, Stanford University}
\author{Amir H. Safavi-Naeini} 
\email[]{safavi@stanford.edu}
\affiliation{Department of Applied Physics \& E.L. Ginzton Laboratory, Stanford University}

\date{\today}

\begin{abstract}
We introduce millimeter-wave silicon photonic crystal cavities as a versatile platform for the perturbative sensing of nanoscale materials. This dielectric-based platform is compatible with strong magnetic fields, opening avenues for studying quantum materials in extreme environments where superconducting cavities cannot operate. To establish the platform's performance, we cryogenically characterize a silicon photonic crystal cavity at 4.3 K, achieving a total quality factor exceeding $10^5$ for a 96 GHz mode. As a proof-of-concept for its sensing capabilities, we position a hexagonal boron nitride-multilayer graphene (hBN-MLG) heterostructure at an electric-field antinode of the cavity and measure the perturbative response at room temperature. The heterostructure induces a significant change in the cavity's resonance, from which we extract a total sample conductivity of approximately $5.1\times10^6$~S/m. These results establish silicon photonic crystal cavities as a promising platform for sensitive, on-chip spectroscopy of nanoscale materials at millimeter-wave frequencies.
\end{abstract}


\maketitle
\stoptoc
\section{Introduction}
\label{sec:intro}
The millimeter-wave (mm-wave) regime (30-300 GHz) bridges the microwave and optical frequencies, providing unique opportunities for both fundamental science and emerging technologies~\cite{kudelin:2024:photonicchipbased, sun:2024:integratedoptical, zhang:2025:opticallyaccessiblea, benea-chelmus:2019:electricfield, hauer:2021:quantumoptomechanics,hauer:2023:nonlinearsideband, xie:2023:subterahertzelectromechanics, xie:2024:subterahertzoptomechanics,legaie:2024:millimeterwaveatomic,multani:2024:quantumlimits, anferov:2025:millimeterwavesuperconducting, multani:2026:integratedmillimeterwave}. This frequency range enables resonant access to elementary excitations and collective modes in quantum materials, rotational modes of small polar molecules, and the structural vibrations of large biomolecules~\cite{jepsen:2011:terahertzspectroscopy, kampfrath:2013:resonantnonresonant, joyce:2016:reviewelectrical, dhillon:2017:2017terahertz, spies:2020:terahertzspectroscopy, bera:2021:reviewrecent}.
Ultrafast terahertz (THz) time-domain spectroscopy is an indispensable tool in this landscape, yet its free-space architecture and optical gating can be challenging to implement in cryogenic or high-field environments. Moreover, typical detection mechanisms in time-domain THz spectroscopy, such as photoconductive antennas or electro-optic sampling, often suffer from limited sensitivity when probing very low-energy excitations or subtle electronic states~\cite{cocker:2021:nanoscaleterahertz,withayachumnankul2014limitation}.

Meanwhile, advances in mm-wave and THz integrated photonics have opened new avenues for miniaturizing future telecommunication and sensing systems~\cite{headland:2023:terahertzintegration, rajabali:2023:presentfuture, lampert:2025:photonicsintegratedterahertz}. The small wavelength of light at these frequencies and the excellent properties of silicon at mm-wave and THz wavelengths enable low-cost manufacturing approaches based on silicon micromachining.

\begin{figure}
	\centering
	\includegraphics[]{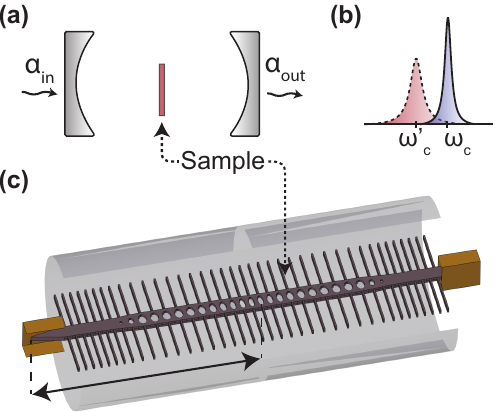}
	\caption{\textbf{Sensing via cavity perturbation using silicon photonic crystal cavities}. (a) Diagram of a sample placed inside a Fabry-Pérot cavity. (b) Diagram of the transmission through the cavity $|\alpha_\mathrm{out}/\alpha_\mathrm{in}|^2$ around one of its modes, measured before ($\omega_\text{c}$) and after the sample has been placed inside the cavity ($\omega_\text{c}'$). The cavity linewidth can also change due to the introduction of the sample $\kappa_\text{i}\to\kappa_\text{i}'$ (blue shading to red shading). (c) Rendering of a millimeter-wave photonic crystal cavity. To maximize the cavity's perturbative response, the sample is introduced at an electric field antinode of the photonic crystal cavity fundamental mode. The solid arrow indicates half of the device's length, 18.74 mm.}
	\label{fig:fig1}
\end{figure}
In this Letter, we present a silicon mm-wave device for the perturbative sensing of material properties at $\sim$100 GHz. Although high-$Q$ silicon photonic crystal cavities have been used for biosensing at frequencies ranging from mm-wave to optical~\cite{omer:2020:wgmbasedsensing, zhao:2022:sensitiverobust, zhao:2022:photoniccrystal,
hosseinifarahabadi:2022:subterahertzsiliconbased, altug:2022:advancesapplications}, their application to nanoscale materials has not yet been explored. As an initial demonstration, we measured the perturbative effects of thin multilayer graphene flakes (MLG) on the modes of a silicon photonic crystal cavity. Importantly, we measured the cryogenic performance of this cavity architecture for the first time, observing quality factors exceeding $10^5$ in the W-band (75-110 GHz)~\cite{lin:1996:highqphotonic, otter:2014:100ghz, hanham:2017:ledswitchablehighq, zhao:2022:sensitiverobust, zhao:2022:photoniccrystal, salek:2023:highq100}. Unlike traditional free-space, time-domain approaches that measure averaged material responses, perturbative sensing with a silicon photonic crystal cavity enables frequency-domain detection with inherent sub-wavelength field confinement. As depicted in~\cref{fig:fig1}, in perturbation-based sensing, changes to the cavity resonance frequency, $\Delta\omega = \omega_\text{c}'-\omega_\text{c}$, and internal linewidth, $\Delta\kappa_\text{i} = \kappa_\text{i}' -\kappa_\text{i}$, directly encode the complex permittivity, $\tilde\varepsilon = \varepsilon'-j\varepsilon''$. Consequently, a high quality factor, $Q$, is critical to detect small changes in $\tilde\varepsilon$. 

\section{Photonic Crystal Cavity Design \& Fabrication}
\label{sec:design}
\begin{figure}
	\centering
	\includegraphics[]{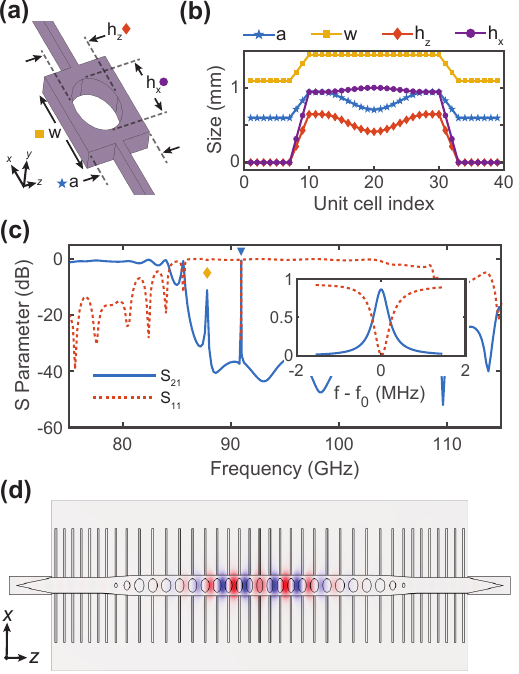}
	\caption{\textbf{Photonic crystal cavity design and COMSOL simulation results}.~(a) Geometry of the nominal unit cell, with critical dimensions labeled with parameters and marker styles corresponding to the following subfigure. (b) Distribution of the unit cell parameters along the device, showing the waveguide region at the two ends, the linear taper between the waveguide and the mirror cells, and the cubic interpolation between the mirror cell and the defect cell. (c) Frequency domain simulation of the full device, coupled to WR10 waveguides. We plot the transmission ($S_{21}$) and reflection ($S_{11}$) of the cavity. The fundamental mode is indicated by a blue triangle and the first-order mode is indicated by a yellow diamond. Inset: plot of the fundamental cavity mode, with a 90.958 GHz center frequency and a total linewidth of roughly 450 kHz (dominated by external coupling). (d) The electric field distribution $E_x$ of the fundamental mode (TE0) of the full device.} 
	\label{fig:fig2}
\end{figure}

Our millimeter-wave silicon photonic crystal cavity adapts optical frequency photonic crystal cavity design principles to achieve photonic confinement~\cite{chan:2012:optimizedoptomechanical}. The design process, summarized in~\cref{fig:fig2}, involves first engineering a photonic bandgap through periodic index modulation, then introducing an optimized defect for mode confinement, and finally implementing linear tapers to interface with WR10 rectangular waveguides (see~\cref{sec:proploss} and~\cref{fig:PropLoss}a).

The periodic mirror cells create a photonic bandgap by coupling forward- and backward-propagating transverse-electric (TE) modes. Using finite-element simulations (COMSOL), we engineer a bandgap centered at 97.29 GHz with a 22.86 GHz span (23.5\% fractional bandwidth), as shown in~\cref{fig:fig2}a,b. A resonant cavity is formed by introducing a defect cell, with a smooth cubic interpolation between the mirror and defect geometries~\cite{chan:2012:optimizedoptomechanical}. A genetic algorithm was used to maximize the radiation-limited quality factor of the fundamental defect mode. The fundamental mode is defined as the mode with the fewest longitudinal variations in the electric field confined by the cavity. In this quasi-1D cavity, the resonance frequency decreases as the longitudinal mode number increases, due to the corresponding increase in the effective refractive index. Full-device, frequency-domain simulations predict a fundamental mode at 90.958 GHz with a total linewidth of $\approx450~\text{kHz}$, dominated by external coupling to the WR10 rectangular waveguides (see~\cref{fig:fig2}d). The complete design parameters as a function of unit cell index are shown in~\cref{fig:fig2}c.

We fabricated the photonic crystal cavities from intrinsic silicon ($\rho_\text{Si} \geq 20 \text{ k}\Omega$-cm) using the device geometry determined by the optimization procedure described above. The main features were defined by photolithography, and the silicon was etched using a reactive-ion etch (see~\cref{sec:fab} and the attached device DXF).

\section{Sensing Multilayer graphene}
\label{sec:measurements}
\begin{figure*}
	\centering
	\includegraphics[]{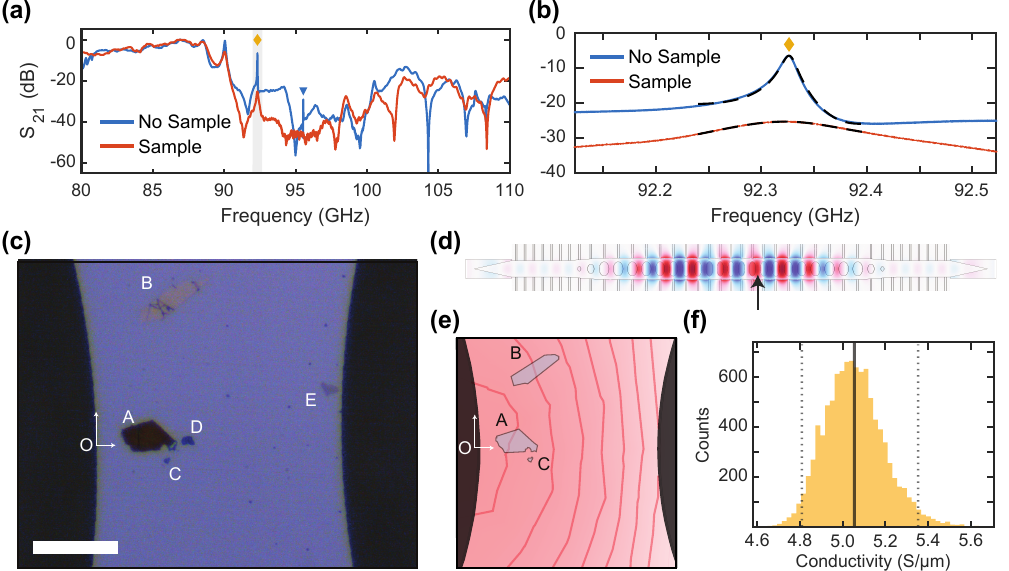}
	\caption{\textbf{Millimeter-wave silicon photonic crystal cavity sensing measurements at room temperature and pressure}.~(a) Normalized transmission through the photonic crystal cavity with (red) and without (blue) the MLG heterostructure.~(b) Close-up of the first harmonic (golden diamond) mode. The black-dash lines represent fits using~\cref{eq:s21}, fitting $\omega_\text{c}, \kappa$ and $t$. The external coupling is determined from fitting the reflection (\cref{eq:s11}).~(c) Micrograph showing the flakes after the dry-transfer process. We see multiple flakes which we label A-E, where D and E are hBN and the rest are MLG. The scale bar indicates 100 \textmu m.~(d) Electric-field distribution, $E_x$, of the first-order mode from a frequency domain simulation. The arrow shows where the sample was transferred.~(e) An illustration of only conductive flakes, with the electric field distribution exported from (d).~(f) The conductivity given by~\cref{eq:cond} of the samples, using the measured sample volume and positional information. The histogram was generated via sampling the positions of flakes A-C from a normal distribution with a standard deviation of 12.5 μm.}
	\label{fig:fig3}
\end{figure*}

As a proof-of-concept, we integrated a two-dimensional quantum material heterostructure directly onto the silicon photonic crystal cavity (\cref{fig:fig3}d,e). We utilized a dry-transfer technique in which hexagonal boron nitride (hBN) serves as the pick-up layer, ensuring an atomically clean interface that encapsulates the MLG (see~\cref{sec:ghBN}). The hBN-MLG stack was aligned to an electric field antinode of the fundamental mode to maximize the perturbative interaction (\cref{fig:fig3}c,d). Following transfer, we verified the heterostructure morphology using microwave impedance microscopy (MIM) \cite{barber2022microwave,ji2025local}, identifying three distinct MLG flakes (A, B, and C) with thicknesses of $t_a = 5$ nm, $t_b = 16$ nm, and $t_c = 5$ nm, alongside adjacent hBN-only regions.

Sensing measurements were performed at room temperature and atmospheric pressure by comparing the transmission spectra of the photonic crystal cavity with and without the MLG heterostructures. The measurements are shown in~\cref{fig:fig3}a. At 100 GHz, these measurements directly probe the low-energy Drude response of 2D conductors such as MLG, providing a complementary window to DC transport and infrared/optical spectroscopy.

The transmission spectrum reveals prominent shifts within the bandgap upon introducing the sample. While the sample perturbs other non-defect resonances within the gap ($>$87 GHz), we focus on the fundamental and first-order defect modes, indicated by a blue triangle and a gold diamond, respectively, shown in~\cref{fig:fig3}a.

In our case, we intended to utilize the fundamental mode for sensing; however, its visibility was lost after the sample transfer due to a low external coupling rate. Therefore, we used the first-order mode for sensing. From first-order perturbation theory (see~\cref{sec:ptheory}), and by writing the imaginary part of the complex permittivity as $\varepsilon'' = \sigma/\omega$, we can relate the sample's conductivity to the unperturbed electric field distribution by
\begin{align}
    \label{eq:cond}
    \sigma = \Delta \kappa_\text{i}  \times \frac{\iiint_V\varepsilon|\mathbf{E}|^2 \mathrm{dV}}{\iiint_{V_\text{sample}} |\mathbf{E}|^2 \mathrm{dV}}.
\end{align}
To determine the conductivity, we measured the change in the internal cavity loss. Since we have multiple MLG flakes, the denominator of~\cref{eq:cond} can be simplified to $\iiint_{V_\text{sample}} |\mathbf{E}|^2 \mathrm{dV} \approx \sum_{i} V_{\text{sample},i} \times |\mathbf{E}_{c,i}|^2$, where $\mathbf{E}_{c,i}$ is the electric field at the center of the $i$-th flake. We determined the volume of our samples by multiplying the thickness measured via MIM by the area measured from an optical micrograph (\cref{fig:fig3}c).

We measured both the reflection and transmission of the photonic crystal cavity with and without the hBN-MLG heterostructure. From input-output theory (see~\cref{sec:lineshape}), the transmission is
\begin{equation}
    \label{eq:s21}
    S_{21}(\Omega) = t - \frac{\kappa_\text{e}}{i(\omega_\text{c}-\Omega)+\kappa/2}.
\end{equation}
In~\cref{eq:s21}, $\omega_\text{c}$ is the mode frequency, $\Omega$ is the excitation frequency, $\kappa$ is the total linewidth, $\kappa_\text{e}$ is the single-sided external coupling rate, and $t$ is a complex-valued crosstalk coefficient that causes the lineshape asymmetry. This crosstalk makes it difficult to separate $\kappa_\text{e}$ and $\kappa_\text{i}$ from the total loss rate. However, reflection measurements do not suffer from the same crosstalk problem. Therefore, we first fit the reflection~\cite{khalil:2012:analysismethod},
\begin{equation}
    \label{eq:s11}
    S_{11}(\Omega) = 1 - \frac{\kappa_\text{e}}{i(\omega_\text{c}-\Omega)+\kappa/2}.
\end{equation}
to robustly determine $\kappa_\text{e}$ before sample transfer, then used this value as a fixed parameter when fitting the transmission spectra. A summary of the fits is presented in~\cref{tab:roomTfits}.

\begin{table}[!ht]
\caption{\textbf{Photonic crystal cavity parameters before and after hBN-MLG sample transfer.} The measured mode parameters of the first harmonic ($\blacklozenge$) and the fundamental ($\blacktriangledown$) with and without the hBN-MLG heterostructure. Note $\kappa = \kappa_\text{i} + 2 \kappa_\text{e}$.}
\label{tab:roomTfits}
\begin{tabular*}{\linewidth}{@{\extracolsep{\fill}}lcccc}
\multirow{4}{*}{\textbf{Parameter}} & \multicolumn{2}{c}{$\blacklozenge$} & \multicolumn{2}{c}{$\blacktriangledown$} \\
\cmidrule(lr){2-3} \cmidrule(lr){4-5} 
& \textbf{Bare} & \textbf{Sample} & \textbf{Bare} & \textbf{Sample} \\
\midrule
$\omega_\text{c}/2\pi$ (GHz) & 92.3268 & 92.3288 & 95.5390 & --- \\
$\kappa/2\pi$ (MHz)   & 14.189 & 143.076 & 6.445 & --- \\ 
$\kappa_\text{e}/2\pi$ (MHz) & 3.881 & 3.881 & 0.0696 & --- \\ 
$\kappa_\text{i}/2\pi$ (MHz) & 6.427 & 135.314 & 6.306 & --- \\ 
\bottomrule
\end{tabular*}
\end{table}

From the change in the internal linewidth of the sensing mode (indicated by the gold diamond in~\cref{fig:fig3}b), we measure perturbative shifts in both the cavity frequency and linewidth of $\Delta\omega_\text{c} \approx 2\pi \cdot 2$ MHz and $\Delta\kappa \approx 2\pi\cdot129$ MHz, respectively (see~\cref{tab:roomTfits}). An increase in the cavity frequency implies that the samples are primarily conductive. From the increase in the internal loss rate and the simulated electric field distribution (\cref{fig:fig3}e), we calculate the total conductivity of the samples to be $\sigma_\text{sample} \approx $ 5.1$\rm \times10^6$ S/m, with a 95\% confidence interval of (4.8, 5.4)$\rm \times10^6$ S/m (see~\cref{fig:fig3}f). This conductivity is consistent with theoretical and experimental measurements of multilayer graphene and graphite thin films~\cite{fang:2015:temperaturethicknessdependent, cai:2009:largearea, mogi:2019:ultimatehigh, pirzado:2016:electricalproperty}. The uncertainty in this estimate is dominated by the uncertainty in the flake positions. We model the position of each flake as $\boldsymbol{X}_i \sim \mathcal{N}(\boldsymbol{\mu}_i, \boldsymbol{\Sigma})$, where $\boldsymbol{\mu}_i$ is the mean position measured from the microscope with respect to the origin $O$ (see~\cref{fig:fig3}c) and $\boldsymbol{\Sigma}$ is the covariance matrix. We assume the covariance matrix is diagonal with the standard deviation of the position fixed at 12.5 \textmu m.

\section{Cryogenic measurements}
\label{sec:cryomeasurements}
To evaluate the potential of this platform for sensing, particularly in low-temperature environments, we characterized the performance of the bare cavity cryogenically (see~\cref{sec:setups} for details on the setup). These measurements were performed before the sample transfer.

\begin{figure}
	\centering
	\includegraphics[]{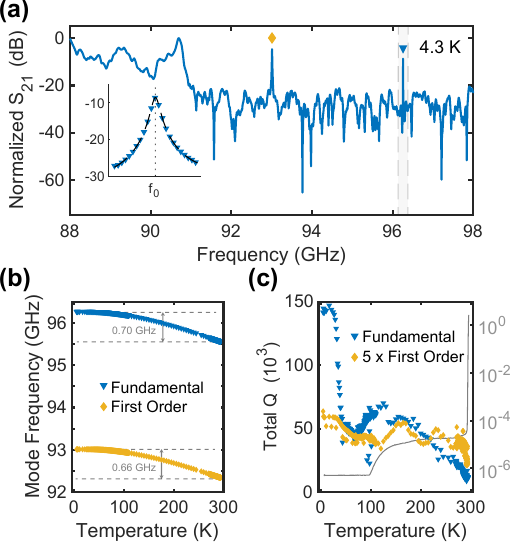}
	\caption{\textbf{Cryogenic measurements of the bare mm-wave silicon photonic crystal cavity}.~(a) Normalized transmission through the setup and photonic crystal cavity at 4.3 K. The shaded region highlights the fundamental mode, where the inset shows the data and the black-dashed line is the corresponding fit. The inferred cavity frequency and linewidth are $\omega_\text{c}\approx 2\pi\cdot 96.261$~GHz and $\kappa \approx 2\pi\cdot 690$~kHz, corresponding to a total quality factor, $Q\approx 1.4\cdot 10^{5}$.~(b) Temperature-dependent cavity frequency of the fundamental (blue triangles) and first harmonic (gold diamonds) modes.~(c) The total quality factor of the fundamental and first harmonic mode, along with a light-gray line corresponding to the \textit{y}-axis on the right side indicating chamber pressure in millibar.}
	\label{fig:fig4}
\end{figure}

At 4.3 K, we measured the total linewidth of the cavity's fundamental mode to be 690 kHz at a resonance frequency of 96.261 GHz, as shown in~\cref{fig:fig4}a. This corresponds to a total (loaded) quality factor of $Q\approx 1.4\cdot10^{5}$ and an internal (unloaded) quality factor of $Q_i \approx 1.7 \cdot10^5$ (see~\cref{tab:cryoFits}). Note that in our cryogenic measurement setup, we were unable to measure reflection (see~\cref{sec:setups}) and thus could not infer the external coupling rate as a function of temperature. As the external coupling rate is determined by the designed cavity geometry, we assume the same $\kappa_\text{e}$ as that measured at room temperature (see~\cref{tab:roomTfits}). At 4.3 K, the first-order mode is extremely overcoupled ($ \kappa_\text{e} \gg \kappa_\text{i} $), making it difficult to separate $\kappa_\text{e}$ and $\kappa_\text{i}$.

\begin{table}[h!]
\caption{\textbf{Photonic crystal bare cavity parameters at 4 K.} The cavity parameters of the first order mode ($\blacklozenge$) and the fundamental mode ($\blacktriangledown$) without the sample, as measured in the cryogenic setup. We assume the same $\kappa_\text{e}$ as at room temperature.}
\label{tab:cryoFits}
\begin{tabular*}{\linewidth}{@{\extracolsep{\fill}}lcccc}
\multirow{4}{*}{\textbf{Parameter}} & \multicolumn{2}{c}{$\blacklozenge$} & \multicolumn{2}{c}{$\blacktriangledown$} \\
\cmidrule(lr){2-3} \cmidrule(lr){4-5} 
& \textbf{298 K} & \textbf{4.3 K} & \textbf{298 K} & \textbf{4.3 K} \\
\midrule
$\omega_\text{c}/2\pi$ (GHz) & 92.3268 & 93.011 & 95.561 & 96.261 \\
$\kappa/2\pi$ (MHz)   & 15.575 & 7.694 & 7.838 & 0.692 \\ 
$\kappa_\text{i}/2\pi$ (MHz) & 7.813  & --- & 7.699 & 0.553 \\ 
\bottomrule
\end{tabular*}
\end{table}

The temperature dependence of the cavity resonance frequencies, shown in~\cref{fig:fig4}b, is due to the changing effective refractive index, $n_\text{eff}$, of the photonic crystal structure. This effective index depends on both the silicon's refractive index and the geometric filling fraction of the structure~\cite{ komma:2012:thermoopticcoefficient, middelmann:2015:thermalexpansion}. For both the fundamental and first-order modes, we observe a total fractional frequency shift of approximately $0.7$\%.

Additionally, both the fundamental and first-order modes experience a strong reduction in internal loss as the temperature decreases, as shown in~\cref{fig:fig4}c and summarized in~\cref{tab:cryoFits}. We attribute this reduction in internal loss to the increased resistivity of intrinsic silicon at low temperatures due to carrier freeze-out~\cite{pires:1990:carrierfreezeout,gehl:2017:operationhighspeed,tobehn-steinhauser:2021:carriermobility}.

Based on silicon's Drude model complex permittivity, the material-limited loss tangent within the W-band is $\tan(\delta) \approx 10^{-4}$ for silicon with a resistivity of $\rho_\text{Si}(298 \text{ K}) = 20 \text{ k}\Omega$-cm (see~\cref{sec:proploss}). At room temperature, our measured internal quality factor of $Q_i\approx1.2\cdot10^{4}$ indicates that the cavity performance is limited by the material loss. If the internal loss at 4.3~K is also material-limited, the measured quality factor implies a bulk silicon resistivity of $\rho_\text{Si}(4.3~\text{K}) \approx 400~\text{k}\Omega\text{-cm}$ (see~\cref{sec:proploss}).

\section{Conclusion}
\label{sec:conclusion}
In this Letter, we have introduced millimeter-wave silicon photonic crystal cavities as a platform for the perturbative sensing of nanoscale materials. We have demonstrated the high performance of this platform through cryogenic measurements, achieving internal quality factors that exceed $10^5$, which is essential for sensitive detection. Our proof-of-concept experiment with an hBN-MLG heterostructure showcased the platform's pronounced sensitivity and its ability to quantify material properties, such as conductivity, at room temperature. While this initial demonstration focuses on a single heterostructure, the agreement between our extracted conductivity and established values validates the technique.

A key advantage of this silicon-based platform is its compatibility with strong magnetic fields, making it a promising tool for studying quantum materials in extreme environments where traditional superconducting cavities are not viable. Although our initial measurements revealed a strong perturbative response that saturated the fundamental mode, we successfully utilized a higher-order mode to extract the sample's conductivity. Future work will focus on optimizing sample placement and cavity design to maintain the system within a trackable perturbative regime.

The inherent flexibility and scalability of this silicon-based platform allow for straightforward adaptation to other frequency ranges, including the terahertz regime. This opens up exciting possibilities for on-chip spectroscopy of a wide range of material excitations, paving the way for the development of integrated sensing applications and hybrid quantum systems. This work establishes silicon photonic crystal cavities as a versatile and powerful platform for exploring the properties of nanoscale materials at millimeter-wave frequencies.



\acknowledgments
K. K. S. M. gratefully acknowledges support from the Natural Sciences and Engineering Research Council of Canada (NSERC) and the insightful discussions with Ali Khalatpour and Geun Ho Ahn. Z. J. acknowledges support from the Panofsky fellowship at the SLAC national laboratory. A portion of this work was supported by the Quantum Science Seed Grant program from Stanford University's Q-Farm initiative. A portion of this work was supported by the US federal government via the U.S. Army Research Office (ARO)/Laboratory for Physical Sciences (LPS) Modular Quantum Gates (ModQ) program (grant no. W911NF-23-1-0254), the Air Force Office of Scientific Research and the Office of Naval Research under award number FA9550-23-1-0338, and the US Department of Energy, Office of Basic Energy Sciences, Division and Materials Science and Engineering through grant no. DE-AC0276SF00515. This material is based upon work supported by the U.S. Department of Energy Office of Science National Quantum Information Science Research Centers as part of the Q-NEXT center. A portion of this work was also funded by Amazon Web Services (AWS). Part of this work was performed at nano@stanford RRID:SCR\_026695. The authors acknowledge that the text was edited for clarity, grammar, and style using generative multi-modal models.

\section*{Data Availability}
All data used in this study are available from the corresponding authors upon reasonable request.

\section*{Competing Interests}
A. H. S.-N. is an Amazon Scholar. The other authors declare no competing interests.

\clearpage
\bibliography{mubsensing}

\clearpage
\onecolumngrid
\begin{center}
	\textbf{\large \textsf{Supplementary information for ``\getTitle"}}
\end{center}
\tableofcontents
\thispagestyle{empty}

\setcounter{section}{0}
\setcounter{equation}{0}
\setcounter{figure}{0}
\setcounter{table}{0}
\setcounter{page}{1}
\renewcommand{\theequation}{S\arabic{equation}}
\renewcommand{\thesection}{S\Roman{section}}
\renewcommand{\thefigure}{S\arabic{figure}}

\resumetoc
\clearpage
\twocolumngrid
\section{Loss estimates of millimeter-wave silicon dielectric waveguides}
\label{sec:proploss}
\begin{figure}[h]
	\centering
	\includegraphics[width=\linewidth]{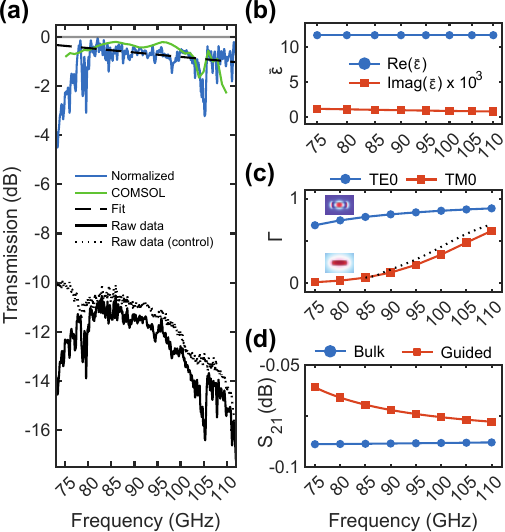}
	\caption{\textbf{Measurements and modeling of silicon millimeter-wave waveguide losses.} (a) Experimental data of 37.487 mm (32.487 mm straight section, 2.5 mm taper sections) tethered silicon waveguide. The black solid line and the black dotted line show the silicon waveguide and a WR10 waveguide of comparable length (25.4 mm), respectively. Normalizing the silicon waveguide to the WR10 waveguide, produces the blue solid line. The green solid line shows a COMSOL frequency domain simulation of the ideal silicon waveguide, corresponding well to the measured data. The black dashed line indicates a fit to the normalized data and has a slope of $-0.02$~dB/GHz and a median value of $-0.67$~dB. (b) Drude model estimates of the complex permittivity, given $\rho_{\text{dc}}= 20$~k\textOmega-cm. The blue circles indicate $\text{Re}(\tilde\varepsilon)$ and the red squares indicate $\text{Im}(\tilde\varepsilon) \times10^{3}$ to place these quantities on the same scale. (c) Energy confinement factor $\Gamma$, for a silicon waveguide with cross-sectional dimensions $1.1$~mm $\times$ 0.38~mm. The blue circles show the TE0 mode, which is shown in the inset (top left) and the red squares show the TM0 mode (bottom left inset). For the TE0 mode, the colors represent $E_{x}$ and for the TM0 the colors represent $H_{x}$. The black dashed line indicates the TE1 mode, which begins to propagate as the frequency increases (wavelength decreases). (d) The insertion loss of a 37.487 mm silicon waveguide inferred from the imaginary component of the Drude-permittivity of bulk silicon (blue circles) and of a guided TE0 mode, given $\rho_\text{dc} =20$~k\textOmega-cm.}
	\label{fig:PropLoss}
\end{figure}

The imaginary component of the complex permittivity, $\tilde\varepsilon$, is a measure of a material's loss. For conductors and semiconductors, the Drude model describes electrical resistance as the stochastic scattering of free-charge carriers. The frequency-dependent Drude complex permittivity can be written as follows~\cite{headland:2023:terahertzintegration}:
\begin{equation}
	\tilde\varepsilon(\omega) = \varepsilon_\infty - \frac{\omega_p^2}{\omega^2+j\omega/\tau},
\end{equation} 
where $\varepsilon_\infty$ is the real-valued relative permittivity in the high-frequency limit, $\omega_p$
is the plasma frequency, and $\tau$ is the scattering relaxation time (the average time interval between carrier scattering events).
Intuitively, if the scattering relaxation time is small, the electrical conductivity is also small, such that $\tau \propto \sigma_\text{dc}$. They can be related through the plasma frequency, $\omega_p$, which is the oscillation frequency of carriers displaced from their equilibrium in the lattice: $\tau = \frac{\sigma_{\text{dc}}}{\varepsilon_0 \omega_p^2}$.
The plasma frequency is related to the free-carrier concentration and the effective mass of the carriers by:
\begin{equation}
	\omega_p = \sqrt{\frac{N q^2}{\varepsilon_0 m_\text{eff}}}.
\end{equation}
The resistivity of a silicon wafer, $\rho_\text{dc} = 1/\sigma_\text{dc}$, varies based on the concentration of impurities within the crystal, which act as free-carrier dopants. For a given resistivity, we can relate the dc conductivity to the concentration of free carriers, $N$, and the carrier mobility, $\mu$, by $\sigma_\text{dc} = q N \mu$. For silicon, the carrier mobility as a function of carrier concentration has been empirically determined to be~\cite{baccarani:1975:electronmobility}:
\begin{equation}
	\mu(N) = \mu_\text{min} + \frac{\mu_\text{max} - \mu_\text{min}}{1 + (N/N_\text{ref})^\chi},
\end{equation}
where $\mu_\text{min} = 92 \text{ cm}^2/(\text{V}\cdot\text{s})$, $\mu_\text{max} = 1360 \text{ cm}^2/(\text{V}\cdot\text{s})$,
$N_\text{ref} = 1.3 \cdot 10^{17} \text{ cm}^{-3}$, and $\chi = 0.91$. For a resistivity of 20 k\textOmega-cm, we can solve the
nonlinear equation for the free-carrier concentration, yielding $N = 2.29\cdot10^{11} \text{ cm}^{-3}$. This is done by combining the previous equation with the equation for dc conductivity. We can then compute the Drude model parameters for our silicon wafer:  $\omega_p = 2\pi \cdot 8.45$ GHz and $\tau = 0.2$ ps, with $\varepsilon_\infty = 11.7$. We use the harmonic mean to calculate the effective mass for charge carriers in silicon, $m_\text{eff} = 3(2/m_t + 1/m_\ell)^{-1}$, where $m_t = 0.1905\cdot m_e$
and $m_\ell = 0.9163\cdot m_e$. The resulting complex permittivity is plotted in~\cref{fig:PropLoss}b.

From the Drude model, we can compute both the bulk and guided insertion loss of the silicon waveguide. The amplitude loss of a propagating wave is given by $\alpha_\text{bulk}^\text{amp} = k_0\cdot\text{Im}(\sqrt{\tilde\varepsilon})$, and the power loss is $\alpha_\text{bulk}^\text{power} = 2 \alpha_\text{bulk}^\text{amp}$, where $k_0 = 2\pi f/c$. To compute the guided loss, we multiply by the geometry- and frequency-dependent confinement factor, $\Gamma$, which is defined as the ratio of the electromagnetic energy within the silicon to the total electromagnetic energy. Thus, the insertion loss of a guided wave is $\alpha_\text{guide}^\text{power} = \alpha_\text{bulk}^\text{power} \times \Gamma(f)$~\cite{headland:2023:terahertzintegration}. The confinement factor and insertion loss are plotted in~\cref{fig:PropLoss}c and~\cref{fig:PropLoss}d, respectively.

To benchmark the performance of our silicon-based waveguides in the mm-wave frequency band, we performed a set of measurements and simulations. The measured and simulated transmission of a 32.487 mm-long straight waveguide in the W-band is shown in~\cref{fig:PropLoss}a. We measured a median loss of 0.7 dB, with a frequency dependence of approximately $-0.02$ dB/GHz, when compared to a 25.4 mm-long rectangular waveguide between 75 and 110 GHz (W-band). We attribute the majority of this loss to the taper, which is corroborated by COMSOL simulations and Drude modeling of the silicon waveguide's propagation loss. From the modeling described above, we expect the waveguide loss to be less than 0.1 dB at these frequencies. Therefore, for a 2.5 mm-long taper (6\% of the total length), we infer a median taper loss of roughly 0.35 dB per facet (corresponding to 92\% transmission) within the W-band.

\section{Modeling asymmetric Lorentzian lineshapes}
\label{sec:lineshape}
Our in-line photonic crystal cavity can be modeled as a two-port system with external coupling rates $\kappa_{e,1}$ and $\kappa_{e,2}$ coupled to a single resonant mode, $\omega_\text{c}$. We probe the cavity at frequency $\Omega$ with a time-harmonic field, $\alpha_\mathrm{in,1}e^{-i\Omega t}$, corresponding to an input power of $P_\mathrm{in,1} = \hbar\Omega|\alpha_\mathrm{in,1}|^2$. The Hamiltonian of the cavity field is:
\begin{equation}
	\hat{H} = \hbar\omega_\text{c}\hat{a}^\dagger\hat{a} - i\hbar\sqrt{\kappa_{e,1}}\Big(\alpha_\mathrm{in, 1}\hat{a}^\dagger e^{-i\Omega t} - \mathrm{h.c.}\Big).
\end{equation}
After applying unitary transformations to rotate into the frame of the probe and to displace the cavity field such that $\hat{a}\to \alpha + \delta\hat{a}$, and neglecting fluctuations $\delta\hat{a}$, the equation of motion for the cavity field can be written as:
\begin{equation}
	\partial_t\alpha = -(i\Delta+\kappa/2)\alpha -\sqrt{\kappa_{e,1}}\alpha_\mathrm{in,1}.
\end{equation}
Here, we define the cavity-probe detuning as $\Delta = \omega_\text{c}-\Omega$ and the total linewidth as $\kappa = \kappa_\text{i} + \kappa_{e,1} + \kappa_{e,2}$. Applying the input-output boundary conditions (with $\alpha_\mathrm{in,2} = 0$), we have:
\begin{equation}
	\alpha_\mathrm{out,2} = \sqrt{\kappa_{e,2}}\alpha,
\end{equation}
from which we can solve for the transmission and reflection transfer functions:
\begin{equation}
	S_{21}(\Omega) = -\frac{\sqrt{\kappa_{e,1}\kappa_{e,2}}}{i\Delta + \kappa/2}.
\end{equation}

Using this transfer function in the presence of asymmetric lineshapes can lead to errors when inferring the internal and external coupling rates. Fabrication imperfections in the rectangular-waveguide-to-silicon transitions can introduce additional transmission paths, producing the asymmetric lineshapes observed in our data. We can model this by rewriting the input-output boundary condition as $\alpha_\mathrm{out,2} = t\alpha_\mathrm{in, 1} + \sqrt{\kappa_{e,2}}\alpha$, which incorporates interference between the cavity-mediated and direct transmission paths. Here, $t = Ae^{-i\varphi}$ is the direct transmission coefficient. As a result, the transmission through the cavity becomes:
\begin{equation}
	S_{21}(\Omega) = t - \frac{\sqrt{\kappa_{e,1}\kappa_{e,2}}}{i(\omega_\text{c}-\Omega) + \kappa/2}.
\end{equation}
Fitting this equation is not a well-posed problem, as $t$ and $\sqrt{\kappa_{e,1}\kappa_{e,2}}$ are correlated. However, we can isolate the external coupling rate by first measuring the reflection:
\begin{equation}
	S_{ii}(\Omega) = 1 - \frac{\kappa_{e,i}}{i(\omega_\text{c}-\Omega) + \kappa/2}.
\end{equation}
By fitting the reflection using analyses of asymmetric resonator transmission developed for superconducting microwave circuits~\cite{khalil:2012:analysismethod}, we can then robustly fit the transmission, providing a complete analysis of our system in the presence of crosstalk.

\section{Notes on cavity perturbation theory}
\label{sec:ptheory}
Cavity perturbation theory is a powerful method for determining the complex permittivity, $\tilde\varepsilon_s$, and permeability, $\tilde\mu_s$, of a sample based on the changes it induces in a resonant cavity. When a small sample is introduced into a cavity, it perturbs the electromagnetic fields, causing a shift in the resonance frequency, $\omega_\text{c}$, and the linewidth, $\kappa$. In this section, we focus on a non-magnetic dielectric perturbation, which is applicable to a wide range of materials at mm-wave frequencies. For a perturbation within a volume $V_\text{sample}$, with $\Delta\varepsilon = \tilde\varepsilon_s - \varepsilon_0$ and $\Delta\mu = 0$, the fractional change in the eigenfrequency for a high-$Q$ cavity ($\tilde\omega_\text{c} = \omega_\text{c} +i\kappa/2\approx \omega_\text{c}$) can be written as:
\begin{equation}
    \frac{\Delta\tilde\omega}{\omega_\text{c}} \approx -\frac{\iiint_{V_\text{sample}} \Delta\varepsilon |\mathbf{E}|^2\mathrm{d}V}{2\iiint_{V} \varepsilon |\mathbf{E}|^2 \mathrm{d}V}.
\end{equation}
Here, $\mathbf{E}$ is the electric field distribution of the unperturbed cavity. By isolating the real and imaginary components, we can write:

\begin{align}
    \varepsilon_s' &= \varepsilon_0 - 2\frac{\Delta\omega_\text{c}}{\omega_{c}}\times\left(\frac{\int_V\varepsilon|\mathbf{E}|^2\mathrm{dV}}{\int_{V_\text{sample}}|\mathbf{E}|^2\mathrm{dV}}\right) \\
    \varepsilon_s'' &= \frac{\Delta\kappa_\text{i}}{\omega_{c}}\times\left(\frac{\int_V\varepsilon|\mathbf{E}|^2\mathrm{dV}}{\int_{V_\text{sample}}|\mathbf{E}|^2\mathrm{dV}}\right)
\end{align}
Similar expressions can be derived for magnetic materials placed at an $\mathbf{H}$-field antinode. When the bare cavity losses are non-negligible, the standard perturbation theory expressions do not hold because they assume a closed system. Derivations that account for perturbations of the cavity fields in an open system can be performed using the quasi-normal mode framework~\cite{yang:2015:simpleanalytical}.

\section{Silicon photonic crystal fabrication}
\label{sec:fab}
The photonic crystal devices were fabricated from a 76.2 mm diameter, 380 \textmu m thick, undoped (\textgreater 20k\textOmega-cm), double-side-polished silicon wafer (University Wafer). Our fabrication process consists of three main steps: defining the hard mask, etching the silicon, and cleaning. To create the etch mask, we first deposit a thin layer of Si$_3$N$_4$, followed by approximately 3 \textmu m of SiO$_2$ on the top side and an additional 600 nm layer on the back side (as an etch stop) using plasma-enhanced chemical vapor deposition (PECVD). We then spin-coat approximately 4.5 \textmu m of SPR 220-3 photoresist on the top side and expose it using a Heidelberg MLA150. Following development in MF26A, we etch the oxide hard mask to prepare the sample for deep reactive-ion etching (DRIE). We secure the sample wafer to a carrier wafer using pump diffusion oil, as both tools only accept 4-inch wafer diameters. After the deep silicon etch, the photonic crystal devices are automatically released and detached from the carrier wafer in an 80$^\circ$C Remover-PG bath for 15 minutes. The devices are then put through an acid cleaning regimen: 10 minutes in a buffered oxide etch (to remove any excess oxide), a 1-minute DI-water rinse (to wash away the BOE), 2 minutes in 30\% potassium hydroxide at 80$^\circ$C (to remove the silicon nitride and smooth sidewall scallops), followed by another DI-water rinse and a final 1-minute buffered oxide etch.

\begin{figure}[h]
	\centering
	\includegraphics[width=\linewidth]{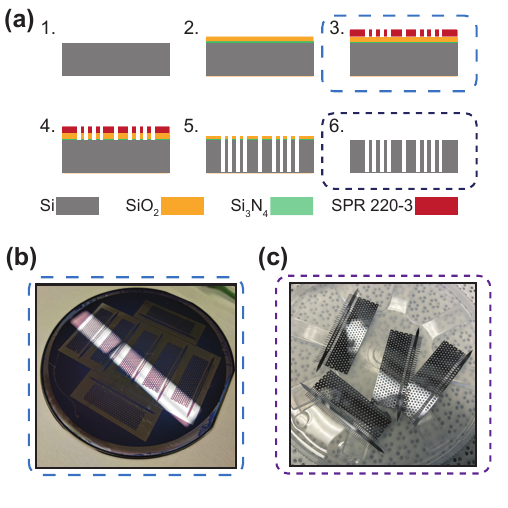}
	\caption{\textbf{Silicon photonic crystal fabrication process.} (a) Step-by-step diagram of our deep-silicon etch fabrication process. (b) Smartphone image of the silicon device wafer on top of the larger carrier wafer (just before oxide etch). (c) Smartphone image of the released photonic crystal devices, taken immediately after an acid cleaning process.}
	\label{fig:fab}
\end{figure}

\section{Graphene sample preparation}
\label{sec:ghBN}
Hexagonal boron nitride (hBN) and graphite flakes were first exfoliated onto SiO$_2$/Si substrates and screened using contrast-enhanced optical microscopy. We then used a poly(bisphenol A) carbonate (PC) film to assemble the heterostructure via a dry pick-up technique. Transferring van der Waals materials onto photonic crystals presents unique challenges compared to bulk substrates. The periodic void structure of the photonic crystal reduces the effective contact area and disrupts the uniformity of the contact force, which can hinder the adhesion of the polymer stamp. To overcome this, we modified the standard protocol by heating the substrate to $200^{\circ}C$ during the release step. This elevated temperature promotes polymer reflow, ensuring robust adhesion of the heterostructure to the suspended silicon network before the PC film is dissolved in chloroform.

\section{Cryogenic experimental setup}
\label{sec:setups}
\begin{figure}[h]
	\centering
	\includegraphics[width=\linewidth]{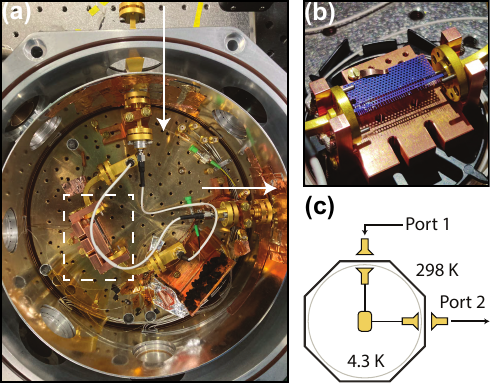}
	\caption{\textbf{Wireless millimeter-wave coupling into a 4K cryostat.} (a) Smartphone image of the cryostat setup. The white arrows indicate the signal path. The white-dashed box shows the photonic crystal cavity we discuss in the main text. (b) Smartphone image of a representative packaged photonic crystal cavity. (c) Minimal diagram of the setup shown in subfigure (a).}
	\label{fig:setup}
\end{figure}

An image of the cryogenic experimental setup is shown in~\cref{fig:setup}. To transmit mm-wave RF signals into the cryostat, we use WR10 horn antennae. Within the cryostat, we use WR10 waveguide-to-coax adapters and coaxial cables to route signals to and from the device under test. The waveguide horns are sensitive to small misalignments within the cryostat, which produced a strong background in the reflection measurement. In our case, this background was strong enough to obscure the photonic crystal cavity's signal.

\clearpage


\end{document}